\hsize=31pc 
\vsize=49pc 
\lineskip=0pt 
\parskip=0pt plus 1pt 
\hfuzz=1pt   
\vfuzz=2pt 
\pretolerance=2500 
\tolerance=5000 
\vbadness=5000 
\hbadness=5000 
\widowpenalty=500 
\clubpenalty=200 
\brokenpenalty=500 
\predisplaypenalty=200 
\voffset=-1pc 
\nopagenumbers      
\catcode`@=11 
\newif\ifams 
\amsfalse %\amstrue si ¤ suivant utilise
% 
%%%%%%%%%%%%%%%%%%%%%%%%%%%%%%%%%%%%%%%%%%%%%%%%%%%%%%%%%%%%% 
%                                                           % 
%  The following section may be commented out and           % 
%  \ifams set to either \amstrue to use the AMS fonts       % 
%  or \amsfalse if they are not available                   % 
%                                                           % 
%%%%%%%%%%%%%%%%%%%%%%%%%%%%%%%%%%%%%%%%%%%%%%%%%%%%%%%%%%%%% 
% 
%\def\Yesreply{Y } 
%\def\Noreply{N } 
%\def\yesreply{y } 
%\def\noreply{n } 
%\newif\ifnotyorn 
%\message{Do you want to use AMSfonts, msam and msbm? Y or N: }% 
%\loop 
%\read-1 to \reply 
%\ifx\reply\yesreply\global\amstrue\notyornfalse 
%\else\ifx\reply\Yesreply\global\amstrue\notyornfalse 
%\else\ifx\reply\noreply\global\amsfalse\notyornfalse 
%\else\ifx\reply\Noreply\global\amsfalse\notyornfalse 
%\else\notyorntrue 
%\message{Please type y or Y  (Yes) or n or N (No)}\fi\fi\fi\fi 
%\ifnotyorn\repeat 
%%%%%%%%%%%%%%%%%%%%%%%%%%%%%%%%%%%%%%%%%%%%%%%%%%%%%%%%%%%% 
% 
\newfam\bdifam 
\newfam\bsyfam 
\newfam\bssfam 
\newfam\msafam 
\newfam\msbfam 
\newif\ifxxpt    
\newif\ifxviipt  
\newif\ifxivpt   
\newif\ifxiipt   
\newif\ifxipt    
\newif\ifxpt     
\newif\ifixpt    
\newif\ifviiipt  
\newif\ifviipt   
\newif\ifvipt    
\newif\ifvpt     
% 
% Headings in 20pt, 17pt or 14pt 
% 
\def\headsize#1#2{\def\headb@seline{#2}% 
                \ifnum#1=20\def\HEAD{twenty}% 
                           \def\smHEAD{twelve}% 
                           \def\vsHEAD{nine}% 
                           \ifxxpt\else\xdef\f@ntsize{\HEAD}% 
                           \def\m@g{4}\def\s@ze{20.74}% 
                           \loadheadfonts\xxpttrue\fi 
                           \ifxiipt\else\xdef\f@ntsize{\smHEAD}% 
                           \def\m@g{1}\def\s@ze{12}% 
                           \loadxiiptfonts\xiipttrue\fi 
                           \ifixpt\else\xdef\f@ntsize{\vsHEAD}% 
                           \def\s@ze{9}% 
                           \loadsmallfonts\ixpttrue\fi 
                      \else 
                \ifnum#1=17\def\HEAD{seventeen}% 
                           \def\smHEAD{eleven}% 
                           \def\vsHEAD{eight}% 
                           \ifxviipt\else\xdef\f@ntsize{\HEAD}% 
                           \def\m@g{3}\def\s@ze{17.28}% 
                           \loadheadfonts\xviipttrue\fi 
                           \ifxipt\else\xdef\f@ntsize{\smHEAD}% 
                           \loadxiptfonts\xipttrue\fi 
                           \ifviiipt\else\xdef\f@ntsize{\vsHEAD}% 
                           \def\s@ze{8}% 
                           \loadsmallfonts\viiipttrue\fi 
                      \else\def\HEAD{fourteen}% 
                           \def\smHEAD{ten}% 
                           \def\vsHEAD{seven}% 
                           \ifxivpt\else\xdef\f@ntsize{\HEAD}% 
                           \def\m@g{2}\def\s@ze{14.4}% 
                           \loadheadfonts\xivpttrue\fi 
                           \ifxpt\else\xdef\f@ntsize{\smHEAD}% 
                           \def\s@ze{10}% 
                           \loadxptfonts\xpttrue\fi 
                           \ifviipt\else\xdef\f@ntsize{\vsHEAD}% 
                           \def\s@ze{7}% 
                           \loadviiptfonts\viipttrue\fi 
                \ifnum#1=14\else 
                \message{Header size should be 20, 17 or 14 point 
                              will now default to 14pt}\fi 
                \fi\fi\headfonts} 
% 
% Text in 12pt, 11pt or 10pt  
% 
\def\textsize#1#2{\def\textb@seline{#2}% 
                 \ifnum#1=12\def\TEXT{twelve}% 
                           \def\smTEXT{eight}% 
                           \def\vsTEXT{six}% 
                           \ifxiipt\else\xdef\f@ntsize{\TEXT}% 
                           \def\m@g{1}\def\s@ze{12}% 
                           \loadxiiptfonts\xiipttrue\fi 
                           \ifviiipt\else\xdef\f@ntsize{\smTEXT}% 
                           \def\s@ze{8}% 
                           \loadsmallfonts\viiipttrue\fi 
                           \ifvipt\else\xdef\f@ntsize{\vsTEXT}% 
                           \def\s@ze{6}% 
                           \loadviptfonts\vipttrue\fi 
                      \else 
                \ifnum#1=11\def\TEXT{eleven}% 
                           \def\smTEXT{seven}% 
                           \def\vsTEXT{five}% 
                           \ifxipt\else\xdef\f@ntsize{\TEXT}% 
                           \def\s@ze{11}% 
                           \loadxiptfonts\xipttrue\fi 
                           \ifviipt\else\xdef\f@ntsize{\smTEXT}% 
                           \loadviiptfonts\viipttrue\fi 
                           \ifvpt\else\xdef\f@ntsize{\vsTEXT}% 
                           \def\s@ze{5}% 
                           \loadvptfonts\vpttrue\fi 
                      \else\def\TEXT{ten}% 
                           \def\smTEXT{seven}% 
                           \def\vsTEXT{five}% 
                           \ifxpt\else\xdef\f@ntsize{\TEXT}% 
                           \loadxptfonts\xpttrue\fi 
                           \ifviipt\else\xdef\f@ntsize{\smTEXT}% 
                           \def\s@ze{7}% 
                           \loadviiptfonts\viipttrue\fi 
                           \ifvpt\else\xdef\f@ntsize{\vsTEXT}% 
                           \def\s@ze{5}% 
                           \loadvptfonts\vpttrue\fi 
                \ifnum#1=10\else 
                \message{Text size should be 12, 11 or 10 point 
                              will now default to 10pt}\fi 
                \fi\fi\textfonts} 
% 
% Small sized material in 10pt, 9pt or 8pt 
% 
\def\smallsize#1#2{\def\smallb@seline{#2}% 
                 \ifnum#1=10\def\SMALL{ten}% 
                           \def\smSMALL{seven}% 
                           \def\vsSMALL{five}% 
                           \ifxpt\else\xdef\f@ntsize{\SMALL}% 
                           \loadxptfonts\xpttrue\fi 
                           \ifviipt\else\xdef\f@ntsize{\smSMALL}% 
                           \def\s@ze{7}% 
                           \loadviiptfonts\viipttrue\fi 
                           \ifvpt\else\xdef\f@ntsize{\vsSMALL}% 
                           \def\s@ze{5}% 
                           \loadvptfonts\vpttrue\fi 
                       \else 
                 \ifnum#1=9\def\SMALL{nine}% 
                           \def\smSMALL{six}% 
                           \def\vsSMALL{five}% 
                           \ifixpt\else\xdef\f@ntsize{\SMALL}% 
                           \def\s@ze{9}% 
                           \loadsmallfonts\ixpttrue\fi 
                           \ifvipt\else\xdef\f@ntsize{\smSMALL}% 
                           \def\s@ze{6}% 
                           \loadviptfonts\vipttrue\fi 
                           \ifvpt\else\xdef\f@ntsize{\vsSMALL}% 
                           \def\s@ze{5}% 
                           \loadvptfonts\vpttrue\fi 
                       \else 
                           \def\SMALL{eight}% 
                           \def\smSMALL{six}% 
                           \def\vsSMALL{five}% 
                           \ifviiipt\else\xdef\f@ntsize{\SMALL}% 
                           \def\s@ze{8}% 
                           \loadsmallfonts\viiipttrue\fi 
                           \ifvipt\else\xdef\f@ntsize{\smSMALL}% 
                           \def\s@ze{6}% 
                           \loadviptfonts\vipttrue\fi 
                           \ifvpt\else\xdef\f@ntsize{\vsSMALL}% 
                           \def\s@ze{5}% 
                           \loadvptfonts\vpttrue\fi 
                 \ifnum#1=8\else\message{Small size should be 10, 9 or  
                            8 point will now default to 8pt}\fi 
                \fi\fi\smallfonts} 
\def\F@nt{\expandafter\font\csname} 
\def\Sk@w{\expandafter\skewchar\csname} 
\def\@nd{\endcsname} 
\def\@step#1{ scaled \magstep#1} 
\def\@half{ scaled \magstephalf} 
\def\@t#1{ at #1pt} 
% 
% For 14, 17 and 20 point fonts use \loadheadfonts 
% 
\def\loadheadfonts{\bigf@nts 
\F@nt \f@ntsize bdi\@nd=cmmib10 \@t{\s@ze}% 
\Sk@w \f@ntsize bdi\@nd='177 
\F@nt \f@ntsize bsy\@nd=cmbsy10 \@t{\s@ze}% 
\Sk@w \f@ntsize bsy\@nd='60 
\F@nt \f@ntsize bss\@nd=cmssbx10 \@t{\s@ze}} 
% 
% For 12 point fonts use \loadxiiptfonts 
% 
\def\loadxiiptfonts{\bigf@nts 
\F@nt \f@ntsize bdi\@nd=cmmib10 \@step{\m@g}% 
\Sk@w \f@ntsize bdi\@nd='177 
\F@nt \f@ntsize bsy\@nd=cmbsy10 \@step{\m@g}% 
\Sk@w \f@ntsize bsy\@nd='60 
\F@nt \f@ntsize bss\@nd=cmssbx10 \@step{\m@g}} 
% 
% For 11 point fonts use \loadxiptfonts 
% 
\def\loadxiptfonts{% 
\font\elevenrm=cmr10 \@half 
\font\eleveni=cmmi10 \@half 
\skewchar\eleveni='177 
\font\elevensy=cmsy10 \@half 
\skewchar\elevensy='60 
\font\elevenex=cmex10 \@half 
\font\elevenit=cmti10 \@half 
\font\elevensl=cmsl10 \@half 
\font\elevenbf=cmbx10 \@half 
\font\eleventt=cmtt10 \@half 
\ifams\font\elevenmsa=msam10 \@half 
\font\elevenmsb=msbm10 \@half\else\fi 
\font\elevenbdi=cmmib10 \@half 
\skewchar\elevenbdi='177 
\font\elevenbsy=cmbsy10 \@half 
\skewchar\elevenbsy='60 
\font\elevenbss=cmssbx10 \@half} 
% 
% For 10 point fonts use \loadxptfonts 
% 
\def\loadxptfonts{% 
\font\tenbdi=cmmib10 
\skewchar\tenbdi='177 
\font\tenbsy=cmbsy10  
\skewchar\tenbsy='60 
\ifams\font\tenmsa=msam10  
\font\tenmsb=msbm10\else\fi 
\font\tenbss=cmssbx10}%  
% 
% For 8 and 9 point fonts use \loadsmallfonts 
% 
\def\loadsmallfonts{\smallf@nts 
\ifams 
\F@nt \f@ntsize ex\@nd=cmex\s@ze 
\else 
\F@nt \f@ntsize ex\@nd=cmex10\fi 
\F@nt \f@ntsize it\@nd=cmti\s@ze 
\F@nt \f@ntsize sl\@nd=cmsl\s@ze 
\F@nt \f@ntsize tt\@nd=cmtt\s@ze} 
% 
% For 7 point fonts use \loadviiptfonts 
% 
\def\loadviiptfonts{% 
\font\sevenit=cmti7 
\font\sevensl=cmsl8 at 7pt 
\ifams\font\sevenmsa=msam7  
\font\sevenmsb=msbm7 
\font\sevenex=cmex7 
\font\sevenbsy=cmbsy7 
\font\sevenbdi=cmmib7\else 
\font\sevenex=cmex10 
\font\sevenbsy=cmbsy10 at 7pt 
\font\sevenbdi=cmmib10 at 7pt\fi 
\skewchar\sevenbsy='60 
\skewchar\sevenbdi='177 
\font\sevenbss=cmssbx10 at 7pt}%  
% 
%  For 6 point fonts use \loadviptfonts 
% 
\def\loadviptfonts{\smallf@nts 
\ifams\font\sixex=cmex7 at 6pt\else 
\font\sixex=cmex10\fi 
\font\sixit=cmti7 at 6pt} 
% 
% For 5 point fonts use \loadvptfonts 
% 
\def\loadvptfonts{% 
\font\fiveit=cmti7 at 5pt 
\ifams\font\fiveex=cmex7 at 5pt 
\font\fivebdi=cmmib5 
\font\fivebsy=cmbsy5 
\font\fivemsa=msam5  
\font\fivemsb=msbm5\else 
\font\fiveex=cmex10 
\font\fivebdi=cmmib10 at 5pt 
\font\fivebsy=cmbsy10 at 5pt\fi 
\skewchar\fivebdi='177 
\skewchar\fivebsy='60 
\font\fivebss=cmssbx10 at 5pt} 
\def\bigf@nts{% 
\F@nt \f@ntsize rm\@nd=cmr10 \@step{\m@g}% 
\F@nt \f@ntsize i\@nd=cmmi10 \@step{\m@g}% 
\Sk@w \f@ntsize i\@nd='177 
\F@nt \f@ntsize sy\@nd=cmsy10 \@step{\m@g}% 
\Sk@w \f@ntsize sy\@nd='60 
\F@nt \f@ntsize ex\@nd=cmex10 \@step{\m@g}% 
\F@nt \f@ntsize it\@nd=cmti10 \@step{\m@g}% 
\F@nt \f@ntsize sl\@nd=cmsl10 \@step{\m@g}% 
\F@nt \f@ntsize bf\@nd=cmbx10 \@step{\m@g}% 
\F@nt \f@ntsize tt\@nd=cmtt10 \@step{\m@g}% 
\ifams 
\F@nt \f@ntsize msa\@nd=msam10 \@step{\m@g}% 
\F@nt \f@ntsize msb\@nd=msbm10 \@step{\m@g}\else\fi} 
\def\smallf@nts{% 
\F@nt \f@ntsize rm\@nd=cmr\s@ze 
\F@nt \f@ntsize i\@nd=cmmi\s@ze  
\Sk@w \f@ntsize i\@nd='177 
\F@nt \f@ntsize sy\@nd=cmsy\s@ze 
\Sk@w \f@ntsize sy\@nd='60 
\F@nt \f@ntsize bf\@nd=cmbx\s@ze  
\ifams 
\F@nt \f@ntsize bdi\@nd=cmmib\s@ze  
\F@nt \f@ntsize bsy\@nd=cmbsy\s@ze  
\F@nt \f@ntsize msa\@nd=msam\s@ze  
\F@nt \f@ntsize msb\@nd=msbm\s@ze 
\else 
\F@nt \f@ntsize bdi\@nd=cmmib10 \@t{\s@ze}%  
\F@nt \f@ntsize bsy\@nd=cmbsy10 \@t{\s@ze}\fi  
\Sk@w \f@ntsize bdi\@nd='177 
\Sk@w \f@ntsize bsy\@nd='60 
\F@nt \f@ntsize bss\@nd=cmssbx10 \@t{\s@ze}}%  
% 
% Fonts for headings  
% 
\def\headfonts{% 
\textfont0=\csname\HEAD rm\@nd         
\scriptfont0=\csname\smHEAD rm\@nd 
\scriptscriptfont0=\csname\vsHEAD rm\@nd 
\def\rm{\fam0\csname\HEAD rm\@nd 
\def\sc{\csname\smHEAD rm\@nd}}% 
\textfont1=\csname\HEAD i\@nd          
\scriptfont1=\csname\smHEAD i\@nd 
\scriptscriptfont1=\csname\vsHEAD i\@nd 
\textfont2=\csname\HEAD sy\@nd         
\scriptfont2=\csname\smHEAD sy\@nd 
\scriptscriptfont2=\csname\vsHEAD sy\@nd 
\textfont3=\csname\HEAD ex\@nd         
\scriptfont3=\csname\smHEAD ex\@nd 
\scriptscriptfont3=\csname\smHEAD ex\@nd 
\textfont\itfam=\csname\HEAD it\@nd    
\scriptfont\itfam=\csname\smHEAD it\@nd 
\scriptscriptfont\itfam=\csname\vsHEAD it\@nd 
\def\it{\fam\itfam\csname\HEAD it\@nd 
\def\sc{\csname\smHEAD it\@nd}}% 
\textfont\slfam=\csname\HEAD sl\@nd    
\def\sl{\fam\slfam\csname\HEAD sl\@nd 
\def\sc{\csname\smHEAD sl\@nd}}% 
\textfont\bffam=\csname\HEAD bf\@nd    
\scriptfont\bffam=\csname\smHEAD bf\@nd 
\scriptscriptfont\bffam=\csname\vsHEAD bf\@nd 
\def\bf{\fam\bffam\csname\HEAD bf\@nd 
\def\sc{\csname\smHEAD bf\@nd}}% 
\textfont\ttfam=\csname\HEAD tt\@nd    
\def\tt{\fam\ttfam\csname\HEAD tt\@nd}% 
\textfont\bdifam=\csname\HEAD bdi\@nd  
\scriptfont\bdifam=\csname\smHEAD bdi\@nd 
\scriptscriptfont\bdifam=\csname\vsHEAD bdi\@nd 
\def\bdi{\fam\bdifam\csname\HEAD bdi\@nd}% 
\textfont\bsyfam=\csname\HEAD bsy\@nd  
\scriptfont\bsyfam=\csname\smHEAD bsy\@nd 
\def\bsy{\fam\bsyfam\csname\HEAD bsy\@nd}% 
\textfont\bssfam=\csname\HEAD bss\@nd  
\scriptfont\bssfam=\csname\smHEAD bss\@nd 
\scriptscriptfont\bssfam=\csname\vsHEAD bss\@nd 
\def\bss{\fam\bssfam\csname\HEAD bss\@nd}% 
\ifams 
\textfont\msafam=\csname\HEAD msa\@nd  
\scriptfont\msafam=\csname\smHEAD msa\@nd 
\scriptscriptfont\msafam=\csname\vsHEAD msa\@nd 
\textfont\msbfam=\csname\HEAD msb\@nd  
\scriptfont\msbfam=\csname\smHEAD msb\@nd 
\scriptscriptfont\msbfam=\csname\vsHEAD msb\@nd 
\else\fi 
\normalbaselineskip=\headb@seline pt% 
\setbox\strutbox=\hbox{\vrule height.7\normalbaselineskip  
depth.3\baselineskip width0pt}% 
\def\sc{\csname\smHEAD rm\@nd}\normalbaselines\bf} 
% 
% Fonts for text 
% 
\def\textfonts{% 
\textfont0=\csname\TEXT rm\@nd         
\scriptfont0=\csname\smTEXT rm\@nd 
\scriptscriptfont0=\csname\vsTEXT rm\@nd 
\def\rm{\fam0\csname\TEXT rm\@nd 
\def\sc{\csname\smTEXT rm\@nd}}% 
\textfont1=\csname\TEXT i\@nd          
\scriptfont1=\csname\smTEXT i\@nd 
\scriptscriptfont1=\csname\vsTEXT i\@nd 
\textfont2=\csname\TEXT sy\@nd         
\scriptfont2=\csname\smTEXT sy\@nd 
\scriptscriptfont2=\csname\vsTEXT sy\@nd 
\textfont3=\csname\TEXT ex\@nd         
\scriptfont3=\csname\smTEXT ex\@nd 
\scriptscriptfont3=\csname\smTEXT ex\@nd 
\textfont\itfam=\csname\TEXT it\@nd    
\scriptfont\itfam=\csname\smTEXT it\@nd 
\scriptscriptfont\itfam=\csname\vsTEXT it\@nd 
\def\it{\fam\itfam\csname\TEXT it\@nd 
\def\sc{\csname\smTEXT it\@nd}}% 
\textfont\slfam=\csname\TEXT sl\@nd    
\def\sl{\fam\slfam\csname\TEXT sl\@nd 
\def\sc{\csname\smTEXT sl\@nd}}% 
\textfont\bffam=\csname\TEXT bf\@nd    
\scriptfont\bffam=\csname\smTEXT bf\@nd 
\scriptscriptfont\bffam=\csname\vsTEXT bf\@nd 
\def\bf{\fam\bffam\csname\TEXT bf\@nd 
\def\sc{\csname\smTEXT bf\@nd}}% 
\textfont\ttfam=\csname\TEXT tt\@nd    
\def\tt{\fam\ttfam\csname\TEXT tt\@nd}% 
\textfont\bdifam=\csname\TEXT bdi\@nd  
\scriptfont\bdifam=\csname\smTEXT bdi\@nd 
\scriptscriptfont\bdifam=\csname\vsTEXT bdi\@nd 
\def\bdi{\fam\bdifam\csname\TEXT bdi\@nd}% 
\textfont\bsyfam=\csname\TEXT bsy\@nd  
\scriptfont\bsyfam=\csname\smTEXT bsy\@nd 
\def\bsy{\fam\bsyfam\csname\TEXT bsy\@nd}% 
\textfont\bssfam=\csname\TEXT bss\@nd  
\scriptfont\bssfam=\csname\smTEXT bss\@nd 
\scriptscriptfont\bssfam=\csname\vsTEXT bss\@nd 
\def\bss{\fam\bssfam\csname\TEXT bss\@nd}% 
\ifams 
\textfont\msafam=\csname\TEXT msa\@nd  
\scriptfont\msafam=\csname\smTEXT msa\@nd 
\scriptscriptfont\msafam=\csname\vsTEXT msa\@nd 
\textfont\msbfam=\csname\TEXT msb\@nd  
\scriptfont\msbfam=\csname\smTEXT msb\@nd 
\scriptscriptfont\msbfam=\csname\vsTEXT msb\@nd 
\else\fi 
\normalbaselineskip=\textb@seline pt 
\setbox\strutbox=\hbox{\vrule height.7\normalbaselineskip  
depth.3\baselineskip width0pt}% 
\everymath{}% 
\def\sc{\csname\smTEXT rm\@nd}\normalbaselines\rm} 
% 
% Fonts for small material (captions, footnotes etc) 
% 
\def\smallfonts{% 
\textfont0=\csname\SMALL rm\@nd         
\scriptfont0=\csname\smSMALL rm\@nd 
\scriptscriptfont0=\csname\vsSMALL rm\@nd 
\def\rm{\fam0\csname\SMALL rm\@nd 
\def\sc{\csname\smSMALL rm\@nd}}% 
\textfont1=\csname\SMALL i\@nd          
\scriptfont1=\csname\smSMALL i\@nd 
\scriptscriptfont1=\csname\vsSMALL i\@nd 
\textfont2=\csname\SMALL sy\@nd         
\scriptfont2=\csname\smSMALL sy\@nd 
\scriptscriptfont2=\csname\vsSMALL sy\@nd 
\textfont3=\csname\SMALL ex\@nd         
\scriptfont3=\csname\smSMALL ex\@nd 
\scriptscriptfont3=\csname\smSMALL ex\@nd 
\textfont\itfam=\csname\SMALL it\@nd    
\scriptfont\itfam=\csname\smSMALL it\@nd 
\scriptscriptfont\itfam=\csname\vsSMALL it\@nd 
\def\it{\fam\itfam\csname\SMALL it\@nd 
\def\sc{\csname\smSMALL it\@nd}}% 
\textfont\slfam=\csname\SMALL sl\@nd    
\def\sl{\fam\slfam\csname\SMALL sl\@nd 
\def\sc{\csname\smSMALL sl\@nd}}% 
\textfont\bffam=\csname\SMALL bf\@nd    
\scriptfont\bffam=\csname\smSMALL bf\@nd 
\scriptscriptfont\bffam=\csname\vsSMALL bf\@nd 
\def\bf{\fam\bffam\csname\SMALL bf\@nd 
\def\sc{\csname\smSMALL bf\@nd}}% 
\textfont\ttfam=\csname\SMALL tt\@nd    
\def\tt{\fam\ttfam\csname\SMALL tt\@nd}% 
\textfont\bdifam=\csname\SMALL bdi\@nd  
\scriptfont\bdifam=\csname\smSMALL bdi\@nd 
\scriptscriptfont\bdifam=\csname\vsSMALL bdi\@nd 
\def\bdi{\fam\bdifam\csname\SMALL bdi\@nd}% 
\textfont\bsyfam=\csname\SMALL bsy\@nd  
\scriptfont\bsyfam=\csname\smSMALL bsy\@nd 
\def\bsy{\fam\bsyfam\csname\SMALL bsy\@nd}% 
\textfont\bssfam=\csname\SMALL bss\@nd  
\scriptfont\bssfam=\csname\smSMALL bss\@nd 
\scriptscriptfont\bssfam=\csname\vsSMALL bss\@nd 
\def\bss{\fam\bssfam\csname\SMALL bss\@nd}% 
\ifams 
\textfont\msafam=\csname\SMALL msa\@nd  
\scriptfont\msafam=\csname\smSMALL msa\@nd 
\scriptscriptfont\msafam=\csname\vsSMALL msa\@nd 
\textfont\msbfam=\csname\SMALL msb\@nd  
\scriptfont\msbfam=\csname\smSMALL msb\@nd 
\scriptscriptfont\msbfam=\csname\vsSMALL msb\@nd 
\else\fi 
\normalbaselineskip=\smallb@seline pt% 
\setbox\strutbox=\hbox{\vrule height.7\normalbaselineskip  
depth.3\baselineskip width0pt}% 
\everymath{}% 
\def\sc{\csname\smSMALL rm\@nd}\normalbaselines\rm}% 
\everydisplay{\indenteddisplay 
   \gdef\labeltype{\eqlabel}}% 
% 
%%%%%%%%%%%%%%%%%%%%%%%%%%%%%%%%%%%%%%%%%%%%%%%%%%%%%%%%%%% 
%                                                         % 
%  Macros to define extra maths symbols                   % 
%                                                         % 
%%%%%%%%%%%%%%%%%%%%%%%%%%%%%%%%%%%%%%%%%%%%%%%%%%%%%%%%%%% 
% 
\def\hexnumber@#1{\ifcase#1 0\or 1\or 2\or 3\or 4\or 5\or 6\or 7\or 8\or 
 9\or A\or B\or C\or D\or E\or F\fi} 
\edef\bffam@{\hexnumber@\bffam} 
\edef\bdifam@{\hexnumber@\bdifam} 
\edef\bsyfam@{\hexnumber@\bsyfam} 
\def\undefine#1{\let#1\undefined} 
\def\newsymbol#1#2#3#4#5{\let\next@\relax 
 \ifnum#2=\thr@@\let\next@\bdifam@\else 
 \ifams 
 \ifnum#2=\@ne\let\next@\msafam@\else 
 \ifnum#2=\tw@\let\next@\msbfam@\fi\fi 
 \fi\fi 
 \mathchardef#1="#3\next@#4#5} 
\def\mathhexbox@#1#2#3{\relax 
 \ifmmode\mathpalette{}{\m@th\mathchar"#1#2#3}% 
 \else\leavevmode\hbox{$\m@th\mathchar"#1#2#3$}\fi} 

\def\bi#1{{\fam\bdifam\relax#1}} 
% 
% If file amsmacro is not in current directory 
% or somewhere with set path add path before 
% file name in following line 
% 
\ifams\input amsmacro\fi 
% 
% Bold italic Greek characters 
% 
\newsymbol\bitGamma 3000 
\newsymbol\bitDelta 3001 
\newsymbol\bitTheta 3002 
\newsymbol\bitLambda 3003 
\newsymbol\bitXi 3004 
\newsymbol\bitPi 3005 
\newsymbol\bitSigma 3006 
\newsymbol\bitUpsilon 3007 
\newsymbol\bitPhi 3008 
\newsymbol\bitPsi 3009 
\newsymbol\bitOmega 300A 
\newsymbol\balpha 300B 
\newsymbol\bbeta 300C 
\newsymbol\bgamma 300D 
\newsymbol\bdelta 300E 
\newsymbol\bepsilon 300F 
\newsymbol\bzeta 3010 
\newsymbol\bfeta 3011 
\newsymbol\btheta 3012 
\newsymbol\biota 3013 
\newsymbol\bkappa 3014 
\newsymbol\blambda 3015 
\newsymbol\bmu 3016 
\newsymbol\bnu 3017 
\newsymbol\bxi 3018 
\newsymbol\bpi 3019 
\newsymbol\brho 301A 
\newsymbol\bsigma 301B 
\newsymbol\btau 301C 
\newsymbol\bupsilon 301D 
\newsymbol\bphi 301E 
\newsymbol\bchi 301F 
\newsymbol\bpsi 3020 
\newsymbol\bomega 3021 
\newsymbol\bvarepsilon 3022 
\newsymbol\bvartheta 3023 
\newsymbol\bvaromega 3024 
\newsymbol\bvarrho 3025 
\newsymbol\bvarzeta 3026 
\newsymbol\bvarphi 3027 
\newsymbol\bpartial 3040 
\newsymbol\bell 3060 
\newsymbol\bimath 307B 
\newsymbol\bjmath 307C 
\mathchardef\binfty "0\bsyfam@31 
\mathchardef\bnabla "0\bsyfam@72 
\mathchardef\bdot "2\bsyfam@01 
\mathchardef\bGamma "0\bffam@00 
\mathchardef\bDelta "0\bffam@01 
\mathchardef\bTheta "0\bffam@02 
\mathchardef\bLambda "0\bffam@03 
\mathchardef\bXi "0\bffam@04 
\mathchardef\bPi "0\bffam@05 
\mathchardef\bSigma "0\bffam@06 
\mathchardef\bUpsilon "0\bffam@07 
\mathchardef\bPhi "0\bffam@08 
\mathchardef\bPsi "0\bffam@09 
\mathchardef\bOmega "0\bffam@0A 
\mathchardef\itGamma "0100 
\mathchardef\itDelta "0101 
\mathchardef\itTheta "0102 
\mathchardef\itLambda "0103 
\mathchardef\itXi "0104 
\mathchardef\itPi "0105 
\mathchardef\itSigma "0106 
\mathchardef\itUpsilon "0107 
\mathchardef\itPhi "0108 
\mathchardef\itPsi "0109 
\mathchardef\itOmega "010A 
\mathchardef\Gamma "0000 
\mathchardef\Delta "0001 
\mathchardef\Theta "0002 
\mathchardef\Lambda "0003 
\mathchardef\Xi "0004 
\mathchardef\Pi "0005 
\mathchardef\Sigma "0006 
\mathchardef\Upsilon "0007 
\mathchardef\Phi "0008 
\mathchardef\Psi "0009 
\mathchardef\Omega "000A 
% 
% Counter definitions 
% 
\newcount\firstpage  \firstpage=1  % start page no 
\newcount\jnl                      % journal no 
\newcount\secno                    % section number 
\newcount\subno                    % number of subsection 
\newcount\subsubno                 % number of subsubsection 
\newcount\appno                    % appendix number 
\newcount\tabno                    % table number 
\newcount\figno                    % figure number 
\newcount\countno                  % equation numbers 
\newcount\refno                    % reference number 
\newcount\eqlett     \eqlett=97    % equation letter 
\newif\ifletter 
\newif\ifwide 
\newif\ifnotfull 
\newif\ifaligned 
\newif\ifnumbysec   
\newif\ifappendix 
\newif\ifnumapp 
\newif\ifssf 
\newif\ifppt 
\newdimen\t@bwidth 
\newdimen\c@pwidth 
\newdimen\digitwidth                    %character width 
\newdimen\argwidth                      %argument width 
\newdimen\secindent    \secindent=5pc   %indentation of maths  
\newdimen\textind    \textind=16pt      %indentation of text 
\newdimen\tempval                       %temporary value 
\newskip\beforesecskip 
\def\beforesecspace{\vskip\beforesecskip\relax} 
\newskip\beforesubskip 
\def\beforesubspace{\vskip\beforesubskip\relax} 
\newskip\beforesubsubskip 
\def\beforesubsubspace{\vskip\beforesubsubskip\relax} 
\newskip\secskip 
\def\secspace{\vskip\secskip\relax} 
\newskip\subskip 
\def\subspace{\vskip\subskip\relax} 
\newskip\insertskip 
\def\insertspace{\vskip\insertskip\relax} 
\def\sp@ce{\ifx\next*\let\next=\@ssf 
               \else\let\next=\@nossf\fi\next} 
\def\@ssf#1{\nobreak\secspace\global\ssftrue\nobreak} 
\def\@nossf{\nobreak\secspace\nobreak\noindent\ignorespaces} 
\def\subsp@ce{\ifx\next*\let\next=\@sssf 
               \else\let\next=\@nosssf\fi\next} 
\def\@sssf#1{\nobreak\subspace\global\ssftrue\nobreak} 
\def\@nosssf{\nobreak\subspace\nobreak\noindent\ignorespaces} 
\beforesecskip=24pt plus12pt minus8pt 
\beforesubskip=12pt plus6pt minus4pt 
\beforesubsubskip=12pt plus6pt minus4pt 
\secskip=12pt plus 2pt minus 2pt 
\subskip=6pt plus3pt minus2pt 
\insertskip=18pt plus6pt minus6pt% 
\fontdimen16\tensy=2.7pt 
\fontdimen17\tensy=2.7pt 
% 
% Labels etc for cross referencing macros 
% 
\def\eqlabel{(\ifappendix\applett 
               \ifnumbysec\ifnum\secno>0 \the\secno\fi.\fi 
               \else\ifnumbysec\the\secno.\fi\fi\the\countno)} 
\def\seclabel{\ifappendix\ifnumapp\else\applett\fi 
    \ifnum\secno>0 \the\secno 
    \ifnumbysec\ifnum\subno>0.\the\subno\fi\fi\fi 
    \else\the\secno\fi\ifnum\subno>0.\the\subno 
         \ifnum\subsubno>0.\the\subsubno\fi\fi} 
\def\tablabel{\ifappendix\applett\fi\the\tabno} 
\def\figlabel{\ifappendix\applett\fi\the\figno} 
\def\gac{\global\advance\countno by 1} 
% 
% Redefinition of footnote macros to lose rule and remove indentation 
% 
 
\def\vfootnote#1{\insert\footins\bgroup 
\interlinepenalty=\interfootnotelinepenalty 
\splittopskip=\ht\strutbox % top baseline for broken footnotes 
\splitmaxdepth=\dp\strutbox \floatingpenalty=20000 
\leftskip=0pt \rightskip=0pt \spaceskip=0pt \xspaceskip=0pt% 
\noindent\smallfonts\rm #1\ \ignorespaces\footstrut\futurelet\next\fo@t} 
% 
% Redefinition of endinsert to give more controllable 
% space around  tables and figures 
% 
\def\endinsert{\egroup 
    \if@mid \dimen@=\ht0 \advance\dimen@ by\dp0 
       \advance\dimen@ by12\p@ \advance\dimen@ by\pagetotal 
       \ifdim\dimen@>\pagegoal \@midfalse\p@gefalse\fi\fi 
    \if@mid \insertspace \box0 \par \ifdim\lastskip<\insertskip 
    \removelastskip \penalty-200 \insertspace \fi 
    \else\insert\topins{\penalty100 
       \splittopskip=0pt \splitmaxdepth=\maxdimen  
       \floatingpenalty=0 
       \ifp@ge \dimen@=\dp0 
       \vbox to\vsize{\unvbox0 \kern-\dimen@}% 
       \else\box0\nobreak\insertspace\fi}\fi\endgroup}    
% 
% special macros for display equations 
% 
% for indentation of turned over lines in mathematics 
% 
\def\ind{\hbox to \secindent{\hfill}} 
% 
% for turned over equals sign to left of maths indent 
% 

% 
% for other signs to left of maths indent 
% 
 
% 
% displayed equation indented  
% 
\def\indeqn#1{\alignedfalse\displ@y\halign{\hbox to \displaywidth 
    {$\ind\@lign\displaystyle##\hfil$}\crcr #1\crcr}} 
% 
% displayed equation indented with alignments 
% 
\def\indalign#1{\alignedtrue\displ@y \tabskip=0pt  
  \halign to\displaywidth{\ind$\@lign\displaystyle{##}$\tabskip=0pt 
    &$\@lign\displaystyle{{}##}$\hfill\tabskip=\centering 
    &\llap{$\@lign\hbox{\rm##}$}\tabskip=0pt\crcr 
    #1\crcr}} 
\def\fl{{\hskip-\secindent}} 
\def\indenteddisplay#1$${\indispl@y{#1 }} 
\def\indispl@y#1{\disptest#1\eqalignno\eqalignno\disptest} 
\def\disptest#1\eqalignno#2\eqalignno#3\disptest{% 
    \ifx#3\eqalignno 
    \indalign#2% 
    \else\indeqn{#1}\fi$$} 
% 
% Roman small caps (if in Roman \sc gives small caps) 
% 
 
% 
% Italic small caps (if in italic \sc gives italic small caps) 
% 
 
% 
% Bold small caps (if in bold \sc gives bold small caps) 
% 
 
% 
% Small caps in maths 
% 
 
% 
% Miscellaneous definitions 
% 

\def\ns{\noalign{\vskip-3pt}}

% 
 
% 
% Bold h bar 
% 
\def\bhbar{\rlap{\kern1pt\raise.4ex\hbox{\bf\char'40}}\bi{h}} 

\def\d{{\rm d}}

\def\frac#1#2{{#1\over#2}} 
\ifams 
\def\lap{\lesssim} 
\def\gap{\gtrsim}

\let\leq=\leqslant

\else

\def\gap{\;\lower3pt\hbox{$\buildrel > \over \sim$}\;}% 
\def\lap{\;\lower3pt\hbox{$\buildrel < \over \sim$}\;}\fi 
\def\i{{\rm i}} 
\chardef\ii="10 
\def\tqs{\hbox to 25pt{\hfil}}

\def\Bbbone{1\kern-.22em {\rm l}} 
% 
% Primes to display summations and products  
% which also have sub or superscripts 
% 
\def\rp{\raise8pt\hbox{$\scriptstyle\prime$}} 
% 
% then use \sum^{...}_{...}\rp or \prod^{...}_{...}\rp. 
% 
% Shadow brackets 
% 
% Single brackets for normal size only 
% 

% 
% Variable size for display style 
% 
\def\[#1\]{\setbox0=\hbox{$\dsty#1$}\argwidth=\wd0 
    \setbox0=\hbox{$\left[\box0\right]$}\advance\argwidth by -\wd0 
    \left[\kern.3\argwidth\box0\kern.3\argwidth\right]} 
% 
% Variable size for text style 
% 
\def\lsb#1\rsb{\setbox0=\hbox{$#1$}\argwidth=\wd0 
    \setbox0=\hbox{$\left[\box0\right]$}\advance\argwidth by -\wd0 
    \left[\kern.3\argwidth\box0\kern.3\argwidth\right]} 
% 
 
% 
% Square for end of theorems 
% 
 
% 
\def\pt(#1){({\it #1\/})} 
\let\dsty=\displaystyle

% 
% Definition for Nuclear Physics Keyword abstract 
% 
\def\reactions#1{\vskip 12pt plus2pt minus2pt%              
\vbox{\hbox{\kern\secindent\vrule\kern12pt% 
\vbox{\kern0.5pt\vbox{\hsize=24pc\parindent=0pt\smallfonts\rm NUCLEAR  
REACTIONS\strut\quad #1\strut}\kern0.5pt}\kern12pt\vrule}}} 
% 
% Definition for slashed characters 
% 
\def\slashchar#1{\setbox0=\hbox{$#1$}\dimen0=\wd0% 
\setbox1=\hbox{/}\dimen1=\wd1% 
\ifdim\dimen0>\dimen1%                         
\rlap{\hbox to \dimen0{\hfil/\hfil}}#1\else                                         
\rlap{\hbox to \dimen1{\hfil$#1$\hfil}}/\fi} 
% 
% Redefine \textindent for use in \item 
% 
\def\textindent#1{\noindent\hbox to \parindent{#1\hss}\ignorespaces} 
% 
% Symbols and curves for use in figure captions 
% 
\def\opencirc{\raise1pt\hbox{$\scriptstyle{\bigcirc}$}} 
 
\ifams 
\def\opensqr{\hbox{$\square$}} 
 
\def\opentridown{\hbox{$\triangledown$}}

\else 
\def\opensqr{\vbox{\hrule height.4pt\hbox{\vrule width.4pt height3.5pt 
    \kern3.5pt\vrule width.4pt}\hrule height.4pt}} 
 
\def\opentridown{\raise1pt\hbox{$\scriptstyle\bigtriangledown$}}

           %  These produce the 
                   %  equivalent open character 
           %  to be filled in. 
\fi

% 
% Redefinition of \cases 
% 
\def\m@th{\mathsurround=0pt} 
% 
% Displaystyle now used for first term 
% 
\def\cases#1{% 
\left\{\,\vcenter{\normalbaselines\openup1\jot\m@th% 
     \ialign{$\displaystyle##\hfil$&\rm\tqs##\hfil\crcr#1\crcr}}\right.}% 
% 
% Original version of cases now called \oldcases 
% 
\def\oldcases#1{\left\{\,\vcenter{\normalbaselines\m@th 
    \ialign{$##\hfil$&\rm\quad##\hfil\crcr#1\crcr}}\right.} 
% 
% Cases with number at end each line (using automatic numbering) 
% 
\def\numcases#1{\left\{\,\vcenter{\baselineskip=15pt\m@th% 
     \ialign{$\displaystyle##\hfil$&\rm\tqs##\hfil 
     \crcr#1\crcr}}\right.\hfill 
     \vcenter{\baselineskip=15pt\m@th% 
     \ialign{\rlap{$\phantom{\displaystyle##\hfil}$}\tabskip=0pt&\en 
     \rlap{\phantom{##\hfil}}\crcr#1\crcr}}} 
\def\ptnumcases#1{\left\{\,\vcenter{\baselineskip=15pt\m@th% 
     \ialign{$\displaystyle##\hfil$&\rm\tqs##\hfil 
     \crcr#1\crcr}}\right.\hfill 
     \vcenter{\baselineskip=15pt\m@th% 
     \ialign{\rlap{$\phantom{\displaystyle##\hfil}$}\tabskip=0pt&\enpt 
     \rlap{\phantom{##\hfil}}\crcr#1\crcr}}\global\eqlett=97 
     \global\advance\countno by 1} 
% 
% for equation numbers instead of \eqno 
% 
\def\eq(#1){\ifaligned\@mp(#1)\else\hfill\llap{{\rm (#1)}}\fi} 
\def\ceq(#1){\ns\ns\ifaligned\@mp\fi\eq(#1)\cr\ns\ns} 
\def\eqpt(#1#2){\ifaligned\@mp(#1{\it #2\/}) 
                    \else\hfill\llap{{\rm (#1{\it #2\/})}}\fi} 
\let\eqno=\eq 
% 
% Automatic numbering of equations 
% 
\countno=1 
 
\def\aleq{&\rm(\ifappendix\applett 
               \ifnumbysec\ifnum\secno>0 \the\secno\fi.\fi 
               \else\ifnumbysec\the\secno.\fi\fi\the\countno} 
\def\noaleq{\hfill\llap\bgroup\rm(\ifappendix\applett 
               \ifnumbysec\ifnum\secno>0 \the\secno\fi.\fi 
               \else\ifnumbysec\the\secno.\fi\fi\the\countno} 
\def\@mp{&} 
\def\en{\ifaligned\aleq)\else\noaleq)\egroup\fi\gac} 
\def\cen{\ns\ns\ifaligned\@mp\fi\en\cr\ns\ns} 
\def\enpt{\ifaligned\aleq{\it\char\the\eqlett})\else 
    \noaleq{\it\char\the\eqlett})\egroup\fi 
    \global\advance\eqlett by 1} 
\def\endpt{\ifaligned\aleq{\it\char\the\eqlett})\else 
    \noaleq{\it\char\the\eqlett})\egroup\fi 
    \global\eqlett=97\gac} 
% 
% abbreviations for Institute of Physics Publishing journals 
% 

\def\JPA{{\it J. Phys. A: Math. Gen.}} 
        %1968-87 
   %1988 and onwards 
     %1968--1988 
        %1989 and onwards 

           %1975--1988 
     %1989 and onwards 
 
                 %1990 and onwards 

% 
% Other commonly quoted journals 
% 

% 
\headline={\ifodd\pageno{\ifnum\pageno=\firstpage\hfill 
   \else\rrhead\fi}\else\lrhead\fi} 
\def\rrhead{\textfonts\hskip\secindent\it 
    \shorttitle\hfill\rm\folio} 
\def\lrhead{\textfonts\hbox to\secindent{\rm\folio\hss}% 
    \it\aunames\hss} 
\footline={\ifnum\pageno=\firstpage \hfill\textfonts\rm\folio\fi} 
\def\@rticle#1#2{\vglue.5pc 
    {\parindent=\secindent \bf #1\par} 
     \vskip2.5pc 
    {\exhyphenpenalty=10000\hyphenpenalty=10000 
     \baselineskip=18pt\raggedright\noindent 
     \headfonts\bf#2\par}\futurelet\next\sh@rttitle}% 
\def\title#1{\gdef\shorttitle{#1} 
    \vglue4pc{\exhyphenpenalty=10000\hyphenpenalty=10000  
    \baselineskip=18pt  
    \raggedright\parindent=0pt 
    \headfonts\bf#1\par}\futurelet\next\sh@rttitle}  

\def\article#1#2{\gdef\shorttitle{#2}\@rticle{#1}{#2}}  
\def\review#1{\gdef\shorttitle{#1}% 
    \@rticle{REVIEW \ifpbm\else ARTICLE\fi}{#1}} 
\def\topical#1{\gdef\shorttitle{#1}% 
    \@rticle{TOPICAL REVIEW}{#1}} 
\def\comment#1{\gdef\shorttitle{#1}% 
    \@rticle{COMMENT}{#1}} 
\def\note#1{\gdef\shorttitle{#1}% 
    \@rticle{NOTE}{#1}} 
\def\prelim#1{\gdef\shorttitle{#1}% 
    \@rticle{PRELIMINARY COMMUNICATION}{#1}} 
\def\letter#1{\gdef\shorttitle{Letter to the Editor}% 
     \gdef\aunames{Letter to the Editor} 
     \global\lettertrue\ifnum\jnl=7\global\letterfalse\fi 
     \@rticle{LETTER TO THE EDITOR}{#1}} 
\def\sh@rttitle{\ifx\next[\let\next=\sh@rt 
                \else\let\next=\f@ll\fi\next} 
\def\sh@rt[#1]{\gdef\shorttitle{#1}} 
\def\f@ll{} 
\def\author#1{\ifletter\else\gdef\aunames{#1}\fi\vskip1.5pc 
    {\parindent=\secindent   
     \hang\textfonts   
     \ifppt\bf\else\rm\fi#1\par}   
     \ifppt\bigskip\else\smallskip\fi 
     \futurelet\next\@unames} 
\def\@unames{\ifx\next[\let\next=\short@uthor 
                 \else\let\next=\@uthor\fi\next} 
\def\short@uthor[#1]{\gdef\aunames{#1}} 
\def\@uthor{} 
\def\address#1{{\parindent=\secindent 
    \exhyphenpenalty=10000\hyphenpenalty=10000 
\ifppt\textfonts\else\smallfonts\fi\hang\raggedright\rm#1\par}% 
    \ifppt\bigskip\fi} 
\def\jl#1{\global\jnl=#1} 
\jl{0}% 
\def\journal{\ifnum\jnl=1 J. Phys.\ A: Math.\ Gen.\  
        \else\ifnum\jnl=2 J. Phys.\ B: At.\ Mol.\ Opt.\ Phys.\  
        \else\ifnum\jnl=3 J. Phys.:\ Condens. Matter\  
        \else\ifnum\jnl=4 J. Phys.\ G: Nucl.\ Part.\ Phys.\  
        \else\ifnum\jnl=5 Inverse Problems\  
        \else\ifnum\jnl=6 Class. Quantum Grav.\  
        \else\ifnum\jnl=7 Network\  
        \else\ifnum\jnl=8 Nonlinearity\ 
        \else\ifnum\jnl=9 Quantum Opt.\ 
        \else\ifnum\jnl=10 Waves in Random Media\ 
        \else\ifnum\jnl=11 Pure Appl. Opt.\  
        \else\ifnum\jnl=12 Phys. Med. Biol.\ 
        \else\ifnum\jnl=13 Modelling Simulation Mater.\ Sci.\ Eng.\  
        \else\ifnum\jnl=14 Plasma Phys. Control. Fusion\  
        \else\ifnum\jnl=15 Physiol. Meas.\  
        \else\ifnum\jnl=16 Sov.\ Lightwave Commun.\ 
        \else\ifnum\jnl=17 J. Phys.\ D: Appl.\ Phys.\ 
        \else\ifnum\jnl=18 Supercond.\ Sci.\ Technol.\ 
        \else\ifnum\jnl=19 Semicond.\ Sci.\ Technol.\ 
        \else\ifnum\jnl=20 Nanotechnology\ 
        \else\ifnum\jnl=21 Meas.\ Sci.\ Technol.\  
        \else\ifnum\jnl=22 Plasma Sources Sci.\ Technol.\  
        \else\ifnum\jnl=23 Smart Mater.\ Struct.\  
        \else\ifnum\jnl=24 J.\ Micromech.\ Microeng.\ 
   \else Institute of Physics Publishing\  
   \fi\fi\fi\fi\fi\fi\fi\fi\fi\fi\fi\fi\fi\fi\fi 
   \fi\fi\fi\fi\fi\fi\fi\fi\fi} 
\let\abs=\beginabstract 

\let\endabs=\endabstract 
\def\submitted{\ifppt\noindent\textfonts\rm Submitted to \journal\par 
     \bigskip\fi} 
\def\today{\number\day\ \ifcase\month\or 
     January\or February\or March\or April\or May\or June\or 
     July\or August\or September\or October\or November\or 
     December\fi\space \number\year} 
\def\date{\ifppt\noindent\textfonts\rm  
     Date: \today\par\goodbreak\bigskip\fi} 
% 
% Physics Abstracts classification numbers 
% 
\def\pacs#1{\ifppt\noindent\textfonts\rm  
     PACS number(s): #1\par\bigskip\fi} 
% 
 
% 
%%%%%%%%%%%%%%%%%%%%%%%%%%%%%%%%%%%%%%%%%%%%%%%%%%%%%%%%%%%% 
%                                                          % 
%  Sections, subsections, etc                              % 
%                                                          % 
%%%%%%%%%%%%%%%%%%%%%%%%%%%%%%%%%%%%%%%%%%%%%%%%%%%%%%%%%%%% 
% 
\def\section#1{\ifppt\ifnum\secno=0\eject\fi\fi 
    \subno=0\subsubno=0\global\advance\secno by 1 
    \gdef\labeltype{\seclabel}\ifnumbysec\countno=1\fi 
    \goodbreak\beforesecspace\nobreak 
    \noindent{\bf \the\secno. #1}\par\futurelet\next\sp@ce} 
\def\subsection#1{\subsubno=0\global\advance\subno by 1 
     \gdef\labeltype{\seclabel}% 
     \ifssf\else\goodbreak\beforesubspace\fi 
     \global\ssffalse\nobreak 
     \noindent{\it \the\secno.\the\subno. #1\par}% 
     \futurelet\next\subsp@ce} 
\def\subsubsection#1{\global\advance\subsubno by 1 
     \gdef\labeltype{\seclabel}% 
     \ifssf\else\goodbreak\beforesubsubspace\fi 
     \global\ssffalse\nobreak 
     \noindent{\it \the\secno.\the\subno.\the\subsubno. #1}\null.  
     \ignorespaces} 
% 
 
% 
%%%%%%%%%%%%%%%%%%%%%%%%%%%%%%%%%%%%%%%%%%%%%%%%%%%%%%%%%%%% 
%                                                          % 
%  Appendices                                              % 
%                                                          % 
%%%%%%%%%%%%%%%%%%%%%%%%%%%%%%%%%%%%%%%%%%%%%%%%%%%%%%%%%%%% 
% 
\def\numappendix#1{\ifappendix\ifnumbysec\countno=1\fi\else 
    \countno=1\figno=0\tabno=0\fi 
    \subno=0\global\advance\appno by 1 
    \secno=\appno\gdef\applett{A}\gdef\labeltype{\seclabel}% 
    \global\appendixtrue\global\numapptrue 
    \goodbreak\beforesecspace\nobreak 
    \noindent{\bf Appendix \the\appno. #1\par}% 
    \futurelet\next\sp@ce} 
\def\numsubappendix#1{\global\advance\subno by 1\subsubno=0 
    \gdef\labeltype{\seclabel}% 
    \ifssf\else\goodbreak\beforesubspace\fi 
    \global\ssffalse\nobreak 
    \noindent{\it A\the\appno.\the\subno. #1\par}% 
    \futurelet\next\subsp@ce} 
\def\@ppendix#1#2#3{\countno=1\subno=0\subsubno=0\secno=0\figno=0\tabno=0 
    \gdef\applett{#1}\gdef\labeltype{\seclabel}\global\appendixtrue 
    \goodbreak\beforesecspace\nobreak 
    \noindent{\bf Appendix#2#3\par}\futurelet\next\sp@ce} 
\def\Appendix#1{\@ppendix{A}{. }{#1}} 
\def\appendix#1#2{\@ppendix{#1}{ #1. }{#2}} 
\def\App#1{\@ppendix{A}{ }{#1}} 
\def\app{\@ppendix{A}{}{}} 
\def\subappendix#1#2{\global\advance\subno by 1\subsubno=0 
    \gdef\labeltype{\seclabel}% 
    \ifssf\else\goodbreak\beforesubspace\fi 
    \global\ssffalse\nobreak 
    \noindent{\it #1\the\subno. #2\par}% 
    \nobreak\subspace\noindent\ignorespaces} 
% 
%%%%%%%%%%%%%%%%%%%%%%%%%%%%%%%%%%%%%%%%%%%%%%%%%%%%%%%%%%%% 
%                                                          % 
%  Acknowledgments, notes added and foreign abstracts      % 
%                                                          % 
%%%%%%%%%%%%%%%%%%%%%%%%%%%%%%%%%%%%%%%%%%%%%%%%%%%%%%%%%%%% 
% 
\def\@ck#1{\ifletter\bigskip\noindent\ignorespaces\else 
    \goodbreak\beforesecspace\nobreak 
    \noindent{\bf Acknowledgment#1\par}% 
    \nobreak\secspace\noindent\ignorespaces\fi} 
\def\ack{\@ck{s}} 
\def\ackn{\@ck{}} 
\def\n@ip#1{\goodbreak\beforesecspace\nobreak 
    \noindent\smallfonts{\it #1}. \rm\ignorespaces} 
\def\naip{\n@ip{Note added in proof}} 
\def\na{\n@ip{Note added}} 
 
% 
%  \resume and \zus in Physics in Medicine and Biology only 
% 
 
% 
 
% 
%%%%%%%%%%%%%%%%%%%%%%%%%%%%%%%%%%%%%%%%%%%%%%%%%%%%%%%%%%%% 
%                                                          % 
%  Tables                                                  % 
%                                                          % 
%%%%%%%%%%%%%%%%%%%%%%%%%%%%%%%%%%%%%%%%%%%%%%%%%%%%%%%%%%% 
% 
 
% 
 
% 
 
\def\tablecont{\topinsert\global\advance\tabno by -1 
    \tablecaption{(continued)}} 
\def\tablecaption#1{\gdef\labeltype{\tablabel}\global\widefalse 
    \leftskip=\secindent\parindent=0pt 
    \global\advance\tabno by 1 
    \smallfonts{\bf Table \ifappendix\applett\fi\the\tabno.} \rm #1\par 
    \smallskip\futurelet\next\t@b} 
\def\t@b{\ifx\next*\let\next=\widet@b 
             \else\ifx\next[\let\next=\fullwidet@b 
                      \else\let\next=\narrowt@b\fi\fi 
             \next} 
\def\widet@b#1{\global\widetrue\global\notfulltrue 
    \t@bwidth=\hsize\advance\t@bwidth by -\secindent}  
\def\fullwidet@b[#1]{\global\widetrue\global\notfullfalse 
    \leftskip=0pt\t@bwidth=\hsize}                   
\def\narrowt@b{\global\notfulltrue} 
\def\align{\catcode`?=13\ifnotfull\moveright\secindent\fi 
    \vbox\bgroup\halign\ifwide to \t@bwidth\fi 
    \bgroup\strut\tabskip=1.2pc plus1pc minus.5pc} 
\def\endalign{\egroup\egroup\catcode`?=12} 
 
% 
% Use \L{#}, \R{#} and \C{#} to specify left, right or centred 
% columns immediately after \table. For example 
% \align\L{#}&&\L{#}\cr gives the preamble for a table with 
% all columns aligned left, \align\L{#}&\C{#}&\R{#}\cr 
% gives a table with 3 columns, the first aligned left, the second 
% centred and the third aligned right. 
% 

% 
%  Rules for tables \br at top and bottom 
%  \mr to separate headings from entries 
% 

% 
 
% 
% Definitions for centring headings over several columns 
% \centre{4}{Results for helium} will centre 
% Results for helium over four columns 
% \crule{4} will produce a rule centred over four columns 
% to go below a centred heading 
% 

% 
 
\catcode`?=13 
\def\lineup{\setbox0=\hbox{\smallfonts\rm 0}% 
    \digitwidth=\wd0% 
    \def?{\kern\digitwidth}% 
    \def\\{\hbox{$\phantom{-}$}}% 
    \def\-{\llap{$-$}}} 
\catcode`?=12 
% 
% Macros for two parts of a table of equal width side by side 
% \table{caption}[w] 
% \sidetable{first part}{second part} 
% \endtable 
% Use \table preamble for tables of 31picas width 
% 
\def\sidetable#1#2{\hbox{\ifppt\hsize=18pc\t@bwidth=18pc 
                          \else\hsize=15pc\t@bwidth=15pc\fi 
    \parindent=0pt\vtop{\null #1\par}% 
    \ifppt\hskip1.2pc\else\hskip1pc\fi 
    \vtop{\null #2\par}}}  
\def\lstable#1#2{\everypar{}\tempval=\hsize\hsize=\vsize 
    \vsize=\tempval\hoffset=-3pc 
    \global\tabno=#1\gdef\labeltype{\tablabel}% 
    \noindent\smallfonts{\bf Table \ifappendix\applett\fi 
    \the\tabno.} \rm #2\par 
    \smallskip\futurelet\next\t@b} 
\def\inctabno{\global\advance\tabno by 1} 
% 
%%%%%%%%%%%%%%%%%%%%%%%%%%%%%%%%%%%%%%%%%%%%%%%%%%%%%%%%%%%% 
%                                                          % 
%  Figures                                                 % 
%                                                          % 
%%%%%%%%%%%%%%%%%%%%%%%%%%%%%%%%%%%%%%%%%%%%%%%%%%%%%%%%%%%% 
% 
 
% 
 
% 
\def\figure#1{\figc@ption{#1}\bigskip} 
\def\figc@ption#1{\global\advance\figno by 1\gdef\labeltype{\figlabel}% 
   {\parindent=\secindent\smallfonts\hang 
    {\bf Figure \ifappendix\applett\fi\the\figno.} \rm #1\par}} 
% 
%%%%%%%%%%%%%%%%%%%%%%%%%%%%%%%%%%%%%%%%%%%%%%%%%%%%%%%%%%%% 
%                                                          % 
%  Reference lists                                         % 
%                                                          % 
%%%%%%%%%%%%%%%%%%%%%%%%%%%%%%%%%%%%%%%%%%%%%%%%%%%%%%%%%%%% 
% 
\def\refHEAD{\goodbreak\beforesecspace 
     \noindent\textfonts{\bf References}\par 
     \let\ref=\rf 
     \nobreak\smallfonts\rm} 
\def\references{\refHEAD\parindent=0pt 
     \everypar{\hangindent=18pt\hangafter=1 
     \frenchspacing\rm}% 
     \secspace} 
\def\rf#1{\par\noindent\hbox to 21pt{\hss #1\quad}\ignorespaces} 
% 
 
% 
 
% 
% reference to a journal article in numerical system 
% 
\def\numrefjl#1#2#3#4#5{\par\rf{#1}#2 {\it #3 \bf #4} #5\par} 
% 
% reference to a book or report in numerical system 
% 
\def\numrefbk#1#2#3#4{\par\rf{#1}#2 {\it #3} #4\par} 
% 
%%%%%%%%%%%%%%%%%%%%%%%%%%%%%%%%%%%%%%%%%%%%%%%%%%%%%%%%%%%% 
%                                                          % 
%  Theorems, lemmas, etc                                   % 
%                                                          % 
%%%%%%%%%%%%%%%%%%%%%%%%%%%%%%%%%%%%%%%%%%%%%%%%%%%%%%%%%%%% 
% 

% 
% NB \note#1 is used to give a Note (as opposed to a paper or letter) 
% in PMB therefore use commands \notes#1 for numbered Note 
% instead of \note  
% 

% 
\catcode`\@=12 
% 
% Parameter values for `Preprint' style  
% 
 
% 
% Parameter values for `Journal' style  
% 
\def\jnlstyle{\pptfalse\headsize{14}{18}% 
\textsize{10}{12}% 
\smallsize{8}{10} 
\textind=16pt} 
% 
% Parameter values for `Eleven point' style  
% 
 
% 
% Parameter values for `Large size' style  
% 
 
% 
\parindent=\textind 
% 
%\magnification=1200
\def\lpsn#1#2{LPSN-#1-LT#2}
\def\received#1{\insertspace 
     \parindent=\secindent\ifppt\textfonts\else\smallfonts\fi 
     \hang{#1}\rm } 
\headline={\ifodd\pageno{\ifnum\pageno=\firstpage\titlehead
   \else\rrhead\fi}\else\lrhead\fi} 
\def\rrhead{\textfonts\hskip\secindent\it 
    \shorttitle\hfill\rm\folio} 
\def\lrhead{\textfonts\hbox to\secindent{\rm\folio\hss}% 
    \it\aunames\hss} 
\footline={\ifnum\pageno=\firstpage{\smallfonts hep-th/9305119}
\hfil\textfonts\rm\folio\fi}   
\def\titlehead{\smallfonts J. Phys. A: Math. Gen.  {\bf 26} 
(1993) 3131--3141\hfil\lpsn{93}{1}} 

\firstpage=3131
\pageno=3131

\jnlstyle
\jl{1}
\overfullrule=0pt

\title{Marginal extended perturbations in two dimensions
and gap--exponent relations}[Marginal extended perturbations]

\author{L Turban and B Berche}[L Turban and B Berche]
 
\address{Laboratoire de Physique du Solide, URA CNRS 155, 
Universit\'e de Nancy I, BP239, F--54506 
Vand\oe uvre l\`es Nancy Cedex, France}

\received{Received 1 February 1993, in final form 19 April 1993}
\abs
The most general form of a marginal extended perturbation 
in a two--dimensional system is deduced 
from scaling considerations. It includes as particular cases 
extended perturbations decaying either from a surface, a line
or a point for which exact results have been previously obtained.
The first--order corrections to the local exponents, which are 
functions of the amplitude of the defect, are deduced from a 
perturbation expansion of the two--point correlation functions.
Assuming covariance under conformal transformation, the perturbed
system is mapped onto a cylinder. Working in the Hamiltonian 
limit, the first--order corrections to the lowest gaps are 
calculated for the Ising model. The results confirm the validity
of the gap--exponent relations for the perturbed system.
\endabs

%\vglue.5cm
\pacs{05.50+q, 05.70Jk, 64.60.Fr}
\submitted
%\date

\section{Introduction}
Following their application to semi--infinite two--dimensional 
systems [1], the methods of conformal invariance have been used
to study the local critical behaviour near line defects [2--4] and 
star--shaped defects [5] in the two--dimensional Ising model.
Although the symmetries usually considered to be necessary for 
conformal invariance [6, 7] (dilatation, rotation and 
translation invariance) are partially broken by such perturbations,
the properties associated with conformal invariance are preserved
and the spectrum--generating algebra have even been identified 
[8--10].

There are some indications that scaling operators remain covariant
under con\-for\-mal trans\-for\-ma\-tion in the presence of marginal
extended perturbations too. This type of defect, which was originally 
introduced
as a deviation from the bulk coupling strength with a power--law 
decay from a free surface in the {\smallfonts 2D} Ising model [11], has been 
extensively studied in the last years [12--17]. It becomes marginal
when the decay exponent is equal to the scaling dimension of
the coupling (see section 2 below). Then local exponents may 
vary with the perturbation amplitude. When the system is mapped 
onto a strip, provided the perturbation profile is properly
transformed, the gap--exponent relations and the tower--like structure
of the spectrum are preserved [18--20] but the associated
algebra has not yet been determined. Similar results were obtained
with an extended perturbation induced by an internal line defect
[21]. This type of perturbation was first considered in [22] (see
also [23--25]). In the case of a radial extended perturbation [26,
27] or with a decaying surface field [28], the gap--exponent 
relations are also satisfied although the equidistant--level 
structure of the spectrum is lost.

In the present work we consider the most general marginal extended
perturbation in the {\smallfonts 2D} Ising model and show that, up to first
order in the defect amplitude, the gap--exponents relations remain
valid. The form of the perturbation is obtained through scaling
considerations in section 2. The first--order corrections to the
local exponents are deduced from a perturbation calculation of
the two--point correlation functions to logarithmic accuracy
in section 3. This is followed by a determination of the 
transformed profile in the strip geometry and a study of the 
condition for shape invariance under the special conformal 
transformation in section 4. The first--order shifts of the lowest
gaps are calculated and the validity of the gap--exponent relations 
examined in section 5. Specific examples are discussed in section 6.
\section{Scaling considerations}
In a continuum description let the Hamiltonian of a two--dimensional 
inhomogeneous system be written as
$$
-\beta H=-\beta H_c+\int\Delta (\bi r)\phi (\bi r)\d^2r\eqno(1)
$$
where $H_c$ is the critical point Hamiltonian of a conformally 
invariant system, $\Delta (\bi r)$ is an extended perturbation, 
with scaling dimension $y_\phi$, which is conjugate to the local 
operator $\phi (\bi r)$ with bulk scaling dimension $x_\phi =
2-y_\phi$. This perturbation can be written as the product of an 
amplitude $g$ by a shape function ${\cal Z}(\bi r)$ which gives 
the form of the inhomogeneity. It is assumed to
be covariant under rescaling so that, in polar coordinates,
$$
{\cal Z}\left({r\over b},\theta\right) =b^\omega{\cal Z}(r,\theta)
\eqno(2)
$$
where the scale invariance of the angle $\theta$ has been taken into 
account. With $b=r$ equation (2) leads to a power--law behaviour for 
the radial part
$$
{\cal Z}(r,\theta )={f(\theta )\over r^\omega}\eqno(3)
$$
whereas the angular dependence remains arbitrary.

The inhomogeneity $\Delta (\bi r)$ transforms according to
$$
\Delta'(\bi r')=g'{\cal Z}\left({r\over b},\theta \right)=b^{y_\phi}
\Delta (\bi r)=b^{y_\phi-\omega}g{\cal Z}\left({r\over b},\theta
\right)\eqno(4)
$$
As a consequence, the perturbation amplitude scales like [29,30]
$$
g'=b^{y_\phi-\omega}g\eqno(5)
$$
When $\omega>y_\phi$ the perturbation decays strongly enough for the 
extended perturbation to be irrelevant. When $\omega<y_\phi$, on the 
contrary, the amplitude increases under rescaling, the perturbation 
is relevant and the original fixed point is unstable. More 
interesting is the borderline situation where $\omega=y_\phi$. Then 
the extended perturbation is marginal and non--universal 
($g$--dependent) local exponents are expected. These results are
easily generalized in higher dimensions. 
\section{Perturbation theory}
We now specialize to the case where the perturbation term in (1) is 
marginal and involves the energy density operator $\varepsilon 
(\bi r)$. Then
$$
-\beta H=-\beta H_c+g\int{\cal Z}(\bi r)\varepsilon 
(\bi r)\d^2r\eqno(6)
$$
The first--order change in the local critical behaviour can be 
deduced 
from a perturbation expansion of the correlation functions [22]. 
The order parameter two--point correlation function has the 
following expansion in powers of $g$
$$						   
\fl G_{\sigma\sigma}(\bi R, g)=\sum_{n=0}^\infty {g^n\over n!}\int
\langle\langle
\sigma (0)\sigma (\bi R)\varepsilon (\bi r_1)\varepsilon (\bi r_2)
\cdots\varepsilon(\bi r_n)\rangle\rangle \prod_{i=1}^n{\cal Z}(\bi
r_i)\d^2r_i \eqno(7)
$$
where the double brackets denote the irreducible part of a 
multi--point correlation function and $\bi R$ is far from the defect
in the bulk of the system. The $n$--th order contribution 
can be rewritten as
$$
\fl \delta G_{\sigma\sigma}^{(n)}= g^n \int_{r_1<r_2\cdots
<r_n}\langle\langle \sigma (0)\sigma (\bi R)\varepsilon (\bi
r_1)\varepsilon(\bi r_2) \cdots\varepsilon (\bi r_n)\rangle\rangle
\prod_{i=1}^n{\cal Z}(\bi r_i)\d^2r_i \eqno(8)
$$
where the $n!$ in the prefactor disappears due to the ordering
on the $r_i$.
The main contribution to the multiple integral comes from regions 
where operators are grouped pairwise close together. Then the 
operator product expansion can be used to reduce the 
multi--point function. With a perturbation in the bulk of the system 
the following reduction relations are needed
$$
\sigma ({\bi r}_1)\varepsilon({\bi r}_2)\simeq a\sigma({\bi r}_1)
r_{12}^{-x_\varepsilon}\qquad\varepsilon({\bi r}_1)\varepsilon 
({\bi r}_2)\simeq b\varepsilon({\bi r}_1)r_{12}^{-x_\varepsilon}
\eqno(9)
$$	
The coefficient $b$ in the second relation vanishes when the 
system is invariant under duality ($\varepsilon\rightarrow
-\varepsilon$). Then the expansion should be
continued one step further. This occurs for the 
Ising model only in the bulk [31] since the surface case does
not possess the duality symmetry.

Let us first suppose that $r_{01}$ is the smallest distance entering 
the correlation function in (8). Then the first relation in (9) can 
be used and the integration over $r_1$ carried out to give 
$$
a\sigma (0)\int_1^{r_2}\d r_1r_1^{1-\omega -x_\varepsilon}
\int_0^{2\pi}f(\theta_1)\d\theta_1\eqno(10)
$$
Since in the marginal case $\omega =2-x_\varepsilon$, the first 
integral contributes a factor $\ln r_2$. It may be checked that when 
one first contracts other pairs, like $\varepsilon (\bi r_1)
\varepsilon (\bi r_2)$, the result is logarithmically smaller. It 
follows that, to the leading logarithmic order
$$			  
\fl\delta G_{\sigma\sigma}^{(n)}= g^naS_f\int_{r_2<r_3\cdots<r_n}
\ln r_2\langle\langle\sigma (0)\sigma (\bi R)\varepsilon (\bi r_2)\cdots
\varepsilon (\bi r_n)\rangle\rangle \prod_{i=2}^n{\cal Z}(\bi
r_i)\d^2r_i \eqno(11)
$$
where $S_f$ is the angular integral in equation (10). The same 
process can be iterated $n$ times leading to
$$
\delta G_{\sigma\sigma}^{(n)}={1\over n!}\left[gaS_f\ln R\right]^n
\langle\langle\sigma (0)\sigma (\bi R)\rangle\rangle \eqno(12)
$$
Here the $n!$ is restored through the successive integrations
involving increasing powers of a logarithm. Finally the 
perturbation series can be summed giving
$$
G_{\sigma\sigma}(\bi R,g)=G_{\sigma\sigma}(\bi R,0)R^{gaS_f}\eqno(13)
$$
to logarithmic accuracy. The order parameter local scaling dimension 
can be read out of the correlation function. It is indeed 
non--universal and reads
$$
x_\sigma (g)=x_\sigma -ga\int_0^{2\pi}f(\theta )\d\theta +O(g^2)
\eqno(14)
$$
The same method can be applied step by step to the calculation of the 
energy density correlation function using the second reduction 
relation in equation (9). The local scaling dimension is then 
$$
x_\varepsilon (g)=x_\varepsilon -gb\int_0^{2\pi}f(\theta )\d\theta 
+O(g^2)\eqno(15)
$$

Let us now consider the case of an extended perturbation centered on 
the surface of a semi--infinite system [32]. In the direction 
perpendicular to the free surface translation invariance is lost and 
the operator product expansion has to be modified accordingly. When 
the first point is the origin, located on the surface, the structure 
constants in (9) only acquire an angular dependence
$$
\sigma (0)\varepsilon({\bi r})\simeq a(\theta)\sigma(0)r^{-x_
\varepsilon}\qquad\varepsilon (0)\varepsilon({\bi r})\simeq b
(\theta)\varepsilon(0)r^{-x_\varepsilon}\eqno(16)
$$	
where $\theta$ is the polar angle measured from the surface. When 
the second point goes to the surface too, i.e. when $\theta
\simeq r^{-1}$ or $\pi -r^{-1}$, the operator product decays with an 
exponent which is the surface scaling dimension $x_\varepsilon^s=d$ 
[33]. Then one expects quite generally $a(\theta )\sim b(\theta )
\sim (\sin\theta )^{2-x_\varepsilon}$.

For the Ising model, with $x_\varepsilon =1$, one indeed obtains
\footnote{\dag}{The polar angle was measured from the perpendicular 
direction in [32]}[32]
$$
a(\theta )={2\over\pi}\sin\theta\qquad b(\theta )={8\over\pi}\sin
\theta\cos^2\theta\eqno(17)
$$
The bulk values in (9) are $a=(2\pi)^{-1}$ and $b=0$ due to the 
duality symmetry in the {\smallfonts 2D} 
bulk Ising system [31]. In this latter case the expansion is in 
powers of~$g^2$.

The calculation of the two--point correlation functions then proceeds 
as above in the bulk. The main change is introduced by the angular 
dependence of the structure constants and one gets the following 
results for the first--order corrections to the local exponents
$$
\fl x_\sigma^s(g)=x_\sigma^s -g\int_0^{\pi}a(\theta)f(\theta)\d\theta 
+O(g^2)\quad x_\varepsilon^s (g)=x_\varepsilon^s-g\int_0^{\pi}b
(\theta)f(\theta)\d\theta+O(g^2)\eqno(18)
$$
provided the angular integrals are not singular. The same restriction
applies to the bulk results in equations (14) and (15).
\section{Conformal aspects}
In this section we assume that conformal techniques can still be used
in the presence of a general extended marginal perturbation. The 
conformal mapping of the perturbed system, either infinite or 
semi--infinite, onto a strip will be used in the next section to get 
the perturbed gaps, allowing a comparison with the perturbation 
results for the local exponents. 
\subsection{Plane--to--cylinder transformation}
We use the mapping $w=(L/2\pi)\ln z$ of the full plane $z=r\exp(\i
\theta )$ onto a cylinder $w=u+\i v$, $-\infty<u<+\infty$, $0<v<L$ 
[34]. Under this transformation the marginal perturbation $\Delta 
(z)$ with scaling dimension $y_\phi$ is changed into [18] 
$$
\Delta (w)=b(z)^{y_\phi}\Delta (z)\eqno(19)
$$
with a dilatation factor given by
$$
b(z)=\left\vert{\d w\over\d z}\right\vert^{-1} ={2\pi r\over L}
\eqno(20)
$$
The shape function transforms in the same way since $g$ is invariant. 
Using
$$
r=\exp{2\pi u\over L}\qquad \theta ={2\pi v\over L}\eqno(21)
$$
one obtains 
$$
\Delta (u,v)=g\left({2\pi\over L}\right)^{y_\phi}f\left({2\pi v
\over L}\right)\eqno(22)
$$
where translation invariance along the strip, which is directly 
linked to the marginal behaviour, is essential for the conformal 
properties. The perturbation is generally inhomogeneous in the 
transverse direction although translation invariance is preserved in 
the case of a radial perturbation [26, 27]. 
\subsection{Special conformal transformation}
The infinitesimal special conformal transformation also plays an 
important r\^ole. When a system is invariant under this 
transformation, differential equations for correlation functions or 
profiles can be obtained, leading to the asymptotic behaviour
of the first and completely determining the second [1, 35]. 

Let the transformation be written as
$$
z'=z+\epsilon z^2\qquad w=z'=r'\exp (\i\theta')\eqno(23)
$$
Up to $O(\epsilon )$, one obtains 
$$
r'=r+\epsilon r^2\cos\theta\quad\theta'=\theta+\epsilon r\sin\theta
\quad b(r,\theta)=1-2\epsilon r\cos\theta\eqno(24)
$$
Together with (19), this leads to the following transformed 
perturbation
$$
\fl\Delta (r',\theta')=(1-2\epsilon r\cos\theta )^{y_\phi}g{f(\theta )
\over r^{y_\phi}}=\left[ 1-\epsilon r\sin\theta\left( y_\phi\cot
\theta +{\d\ln f\over\d\theta}\right)\right]g'{f(\theta')\over 
r'^{y_\phi}}\eqno(25)
$$
The perturbation is shape--invariant when the coefficient of 
$\epsilon$ vanishes, i. e. when 
$$
\Delta (r,\theta )={g\over\vert r\sin\theta\vert^{y_\phi}}\eqno(26)
$$
This corresponds to an extended perturbation decaying either from 
the surface of a semi--infinite system [11] or from a line 
defect in the bulk [22--25, 21].
Then the correlation functions satisfy the same differential 
equation as in the surface case [1]. Their asymptotic behaviour 
can also be obtained, with defect exponents replacing surface ones.
Transforming the correlation functions via the plane--to--cylinder
mapping, the gap--exponent relation and the tower--like structure
of the spectrum immediately follows, in agreement with exact
results for these perturbations [18, 21].
\section{Gap--exponent relations}
\subsection{Hamiltonian limit and rescaling}
We now consider a temperature--like extended marginal perturbation 
in the {\smallfonts 2D} Ising model at the bulk critical point. On the
cylinder the original perturbation is transformed into (22) with 
$\phi=\varepsilon$ and $y_\varepsilon=1$. In the case of a 
semi--infinite original system $(0<\theta<\pi)$, the transformed 
system is a strip with free boundary conditions ({\smallfonts FBC}) at $v=0$ and 
$v=L/2$ on the cylinder whereas for an infinite original system, 
periodic boundary conditions ({\smallfonts PBC}) have to be taken at $v=L$. 

We work on a square lattice in the extreme anisotropic limit, with 
unperturbed two--spin interactions $K_1\gg 1$ in the time direction 
along the cylinder axis and $K_2\ll1$ in the transverse
one. At the  bulk critical point $K_2=K_1^*$ where $K_1^*$ is a dual
coupling  satisfying $\tanh (K_1^*)=\exp (-2K_1)$. The perturbation is
assumed  to act only on the transverse interactions $K_2$.

In the Hamiltonian limit [36, 7] the system becomes anisotropic
with a correlation length ratio [37]
$$
{\xi_2\over\xi_1}={\cosh(2K_2)\over\cosh(2K_1)}\simeq 2K_1^*\eqno(27)
$$
Isotropy can be restored by rescaling the lattice parameter $a_1$ in 
the time direction to $a_1=2K_1^*$ measured in units of $a_2$ in the 
transverse direction. The row--to--row transfer operator may be written 
as
$$
{\cal T}=\exp (-a_1{\cal H})\eqno(28)
$$
The Ising Hamiltonian then takes the form
$$
{\cal H}=-{1\over 2}\left[\sum_n\sigma_n^z+\sum_n\sigma_n^x
\sigma_{n+1}^x\right]-g{2\pi\over L}\sum_nf\left({2\pi n\over L}
\right)\sigma_n^x\sigma_{n+1}^x\eqno(29)
$$
where the $\sigma$'s are Pauli spin operators. The first part in (29) 
corresponds to a homogeneous system at the bulk critical point, 
properly normalized to give critical excitations with velocity 
$v_s=1$ [38]. The perturbation term keeps the same amplitude as in the 
continuum expression (22) since, lengths being measured in units of 
$a_2$, $\exp(-{\cal H})$ operates a transfer by one unit length in 
the time direction and each perturbed bond in ${\cal H}$ is 
associated with one surface unit on the isotropic rescaled system.
\subsection{Diagonalization of the unperturbed Hamiltonian}
The unperturbed part of the Hamiltonian in (29) is rewritten as a 
quadratic form in fermion operators through a Jordan--Wigner 
transformation [39]
$$
\fl{\cal H}_c(P)=-\sum_{n=1}^N(c_n^+c_n-{1\over 2})-{1\over 2}
\sum_{n=1}^{N-1}(c_n^+-c_n)(c_{n+1}^++c_{n+1})+{1\over 2}P
(c_N^+-c_N)(c_{1}^++c_{1})\cr
\hfil\eqno(30)
$$
where the chain length $N$ is assumed to be even in the following.
For a semi--infinite original system one has to take $N=L/2$ and 
$P=0$ in (30). If the original system 
covers the whole plane then $N=L$ and $P=\pm 1$ is an 
eigenvalue of the parity operator ${\cal P}=\exp(\i\pi\sum_nc_n^+
c_n)$ which commutes with ${\cal H}_c$.

The Hamiltonian is put under diagonal form
$$
{\cal H}_c(P)=\sum_k\varepsilon_k(\eta_k^+\eta_k-{1/2})\eqno(31)
$$
through a canonical transformation [40, 41].
The squares of the excitation energies for the diagonal fermions
in (31), $\varepsilon_k=2\vert\sin{k/2}\vert$,
can be obtained as the solutions of two equivalent eigenvalue 
problems
with normalized eigenvectors, ${\bf\Phi}_k$ 
and ${\bf\Psi}_k$. The wave--vector quantization depends 
on the boundary conditions, i. e. on $P$ :
$$
\fl k(P=+1)=(2p+1){\pi\over N}\quad k(P=-1)=2p{\pi\over N}\quad
-{N\over 2}\leq p\leq{N\over 2}-1\quad({\smallfonts PBC})\eqno(32a)
$$
$$
\fl k(P=0)=(2p+1){\pi\over 2N+1}\qquad 0\leq p\leq N-1\qquad({\smallfonts FBC})
\eqno(32b)
$$
With periodic boundary conditions, the normalized eigenvectors are
given by
$$
\fl\Phi_k(n)=(-1)^n\sqrt{2\over N}\sin kn\quad
\Psi_k(n)=(-1)^n\sqrt{2\over N}\cos k\left(n+{1\over 2}\right)
\quad 0<k<\pi\eqno(33a)
$$
$$
\fl\Phi_k(n)=(-1)^n\sqrt{2\over N}\cos kn\quad
\Psi_k(n)=(-1)^n\sqrt{2\over N}\sin k\left(n+{1\over 2}\right)
\quad -\pi<k<0\eqno(33b)
$$
$$
\fl\Phi_0(n)=-\Psi_0(n)={(-1)^n\over\sqrt{N}}\qquad
\Phi_{-\pi}(n)=-\Psi_{-\pi}(n)={1\over\sqrt{N}}\qquad(P=-1)
\eqno(33c)
$$
whereas one obtains 
$$
\fl\Phi_k(n)=(-1)^n{2\over\sqrt{2N+1}}\cos k\left(n-{1\over 2}\right)
\qquad\Psi_k(n)=(-1)^{n+1}{2\over\sqrt{2N+1}}\sin kn\eqno(34)
$$
with free boundary conditions.
\subsection{Perturbation theory}
Up to first order in the perturbation amplitude, the levels of 
${\cal H}_c$ in (29) are shifted by
$$
E_\alpha(g)-E_\alpha(0)=-g{2\pi\over L}\sum_nf\left({2\pi n\over L}
\right)\langle\alpha\vert\sigma_n^x\sigma_{n+1}^x\vert\alpha\rangle\eqno(35)
$$
The relevant levels in the following are the ground state $\vert 0\rangle$
and the states $\vert\sigma\rangle$ and $\vert\varepsilon\rangle$ of the 
unperturbed
Hamiltonian which are involved in the calculation of the 
lowest gaps.
These states are the lowest ones with nonvanishing matrix elements 
$\langle0\vert\sigma_n^x\vert\sigma\rangle$ and $\langle0\vert\sigma_n^x\sigma_{n+1}^x
\vert\varepsilon\rangle$ with the ground state.

With periodic boundary conditions and $N$ even, $\vert 0\rangle$ is the 
vacuum of ${\cal H}_c(+1)$. $\vert\varepsilon\rangle=\eta_0^+
\eta_1^+\vert 0\rangle$,
which is even, also belongs to the spectrum of ${\cal H}_c(+1)$ and 
contains the two lowest excitations corresponding to $p=0$ and 
$p=-1$ in $k(P=+1)$. $\vert\sigma\rangle=\eta_0^+
\vert 0\rangle_{(-1)}$, which is odd, belongs to the spectrum of 
${\cal H}_c(-1)$. It contains a single excitation with vanishing 
wave--vector corresponding to $p=0$ in $k(P=-1)$. 
As a consequence this state is degenerate with the vacuum 
$\vert 0\rangle_{(-1)}$ of ${\cal H}_c(-1)$.

In the case of free boundary conditions, the ground--state $\vert 0\rangle$ 
is the vacuum of ${\cal H}_c(0)$, $\vert \varepsilon\rangle$ is defined 
as above
with two excitations corresponding to $p=0$ and $p=1$ in $k(P=0)$ 
while $\vert\sigma\rangle=\eta_0^+\vert 0\rangle$ only contains the lowest one.

The matrix elements in (35), like the Hamiltonian, are obtained 
making use of the Jordan--Wigner and canonical transformations 
and read
$$
\fl D_0(n)=\langle0\vert\sigma_n^x\sigma_{n+1}^x\vert 0\rangle=-\sum_k\Psi_k(n)\Phi_k
(n+1)\eqno(36a)
$$
$$
\fl D_\sigma(n)=\langle\sigma\vert\sigma_n^x\sigma_{n+1}^x\vert\sigma\rangle=D_0(n)
+2\Psi_0(n)\Phi_0(n+1)\eqno(36b)
$$
$$
\fl D_\varepsilon(n)=\langle\varepsilon\vert\sigma_n^x\sigma_{n+1}^x
\vert\varepsilon\rangle=
D_0(n)+2\left[\Psi_0(n)\Phi_0(n+1)+\Psi_1(n)\Phi_1(n+1)\right]
\eqno(36c)
$$
In the case of periodic boundary conditions, the final expressions
for the boundary terms ($n=N$, $n+1=1$) have to be multiplied by
$-P$ as in (30) and the eigenvectors ${\bf\Phi}_k$ and ${\bf\Psi}_k$
are those involved in the diagonalization of ${\cal H}_c(P)$ where
$P$ is the parity of the states in the matrix elements. Then, using
(32--34), one obtains
$$
D_0(n)=\left( N\sin{\pi\over 2N}\right)^{-1}\qquad(P=+1)\eqno(37a)
$$
$$
D_\sigma(n)={1\over N}\left(1+\cot{\pi\over 2N}\right)\qquad(P=-1)
\eqno(37b)
$$
$$
D_\varepsilon(n)-D_0(n)=-{4\over N}\sin{\pi\over 2N}\qquad(P=+1)
\eqno(37c)
$$
while for free boundary conditions
$$
D_\sigma(n)-D_0(n)={2\over N}\sin\left({n\pi\over N}\right)+O(N^{-2})
\eqno(38a)
$$
$$
D_\varepsilon(n)-D_0(n)={8\over N}\sin\left({n\pi\over N}\right)
\cos^2\left({n\pi\over N}\right)+O(N^{-2})\eqno(38b)
$$
According to the gap--exponent relation [34], one expects the scaling
dimensions of the operator $\alpha=\sigma$, $\varepsilon$ to be given by
$$
\fl x_\alpha(g)=\lim_{N\rightarrow\infty}{N\over 2\pi}\left[E_\alpha(g)
-E_0(g)\right]\quad x_\alpha^s(g)=\lim_{N\rightarrow\infty}
{N\over\pi}\left[E_\alpha(g)-E_0(g)\right]\eqno(39)
$$
which, together with (35) and (36), leads to the first--order
changes
$$
\fl x_\alpha(g)-x_\alpha(0)=\lim_{N\rightarrow\infty}-g\sum_nf\left(
{2\pi n\over N}\right)\left[D_\alpha(n)-D_0(n)\right]\qquad
({\smallfonts PBC}) \eqno(40a)
$$
$$
\fl x_\alpha^s(g)-x_\alpha^s(0)=\lim_{N\rightarrow\infty}-g\sum_nf
\left({\pi n\over N}\right)\left[D_\alpha(n)-D_0(n)\right]
\qquad ({\smallfonts FBC})\eqno(40b)
$$
In the continuum limit, using (37--38) for large $N$, this transforms
into
$$
x_\sigma(g)={1\over 8}-{g\over 2\pi}\int_0^{2\pi}\d\theta f(\theta)
+O(g^2)\eqno(41a)
$$
$$
x_\varepsilon(g)=1+O(g^2)\eqno(41b) 
$$
for the perturbed scaling dimensions near the source of the 
inhomogeneity in the plane and
$$
x_\sigma^s(g)={1\over 2}-{2g\over\pi}\int_0^{\pi}\d\theta f(\theta)
\sin\theta+O(g^2)\eqno(42a)
$$
$$
x_\varepsilon^s(g)=2-{8g\over\pi}\int_0^{\pi}\d\theta 
f(\theta)\sin\theta\cos^2\theta+O(g^2)\eqno(42b)
$$
in the half--plane, in full agreement with the results of section 3.
\section{Elliptic defects}
As an illustration of the the perturbation results one may
consider an extended defect with elliptic symmetry in the {\smallfonts 2D}
Ising model. The angular dependence
$$
f(\theta)=(\sin^2\theta+\kappa\cos^2\theta)^{-1/2}\qquad 0<\kappa<1
\eqno(43)
$$
interpolates between the line and radial defects corresponding
to $\kappa=0$ and $\kappa=1$, respectively.

In the infinite Ising system, using (41a)
$$
x_\sigma(g)={1\over 8}-{2g\over\pi}{\rm\bf K}(\sqrt{1-\kappa})
+O(g^2)\eqno(44)
$$
where {\bf K} is the complete elliptic integral. The radial defect
result [26, 27, 42], $x_\sigma(g)=1/8-g+O(g^2)$ is recovered in
the limit $\kappa\rightarrow 1$. The correction term displays
a logarithmic divergence in the line defect limit, $\kappa
\rightarrow 0$, which is linked to the singular behaviour of the 
perturbation at $\theta=0$ and $\pi$. Introducing a cut--off, one
gets a jump of the magnetic exponent at $g=0$ and local order
at the bulk critical point for $g>0$ [21, 31, 43].

In a semi--infinite system (43) corresponds to a defect with its
main axis along the surface. Then, using (42)
$$
x_\sigma^s(g)={1\over 2}-{4g\over\pi}{\arcsin\sqrt{1-\kappa}\over
\sqrt{1-\kappa}}+O(g^2)\eqno(45a)
$$
$$
x_\varepsilon^s(g)=2-{8g\over\pi}\left[{\arcsin\sqrt{1-\kappa}\over
(1-\kappa)^{3/2}}-{\sqrt{\kappa}\over 1-\kappa}\right]+O(g^2)
\eqno(45b)
$$
In the surface defect limit, $\kappa=0$, $x_\sigma^s(g)=1/2
-2g+O(g^2)$, $x_\varepsilon^s(g)=2-4g+O(g^2)$ while for the radial
defect, $\kappa\rightarrow 1$, $x_\sigma^s(g)=1/2-4g/\pi+
O(g^2)$, $x_\varepsilon^s(g)=2-16g/(3\pi)+O(g^2)$ in agreement
with know exact results [18, 26].

When the main axis of the defect is perpendicular to the surface,
i. e. with
$$
f(\theta)=(\cos^2\theta+\kappa\sin^2\theta)^{-1/2}\qquad 0<\kappa<1
\eqno(46)
$$
one obtains
$$
x_\sigma^s(g)={1\over 2}-{4g\over\pi\sqrt{1-\kappa}}\ln\left({
\sqrt{1-\kappa}+1\over\sqrt{\kappa}}\right)+O(g^2)\eqno(47a)
$$
$$
x_\varepsilon^s(g)=2-{8g\over\pi}\left[{1\over1-\kappa}-{\kappa\over
(1-\kappa)^{3/2}}\ln\left({\sqrt{1-\kappa}+1\over\sqrt{\kappa}}
\right)\right]+O(g^2)\eqno(47b)
$$
with the same limits as in (45) for the radial defect ($\kappa=1$). 
When $\kappa=0$, i. e. with a line defect perpendicular to the 
surface, $x_\varepsilon^s(g)=2-8g/\pi+O(g^2)$ whereas the 
correction to $x_\sigma^s$ diverges logarithmically. Then one 
expects, as for a bulk extended line defect, a jump of the surface
magnetic exponent at $g=0$ and local order for enhanced couplings.
This could be checked using the techniques of [18, 21].
\section{Conclusion}
Our main result is the extension of the gap--exponent relations
to the case of a general marginal extended perturbation in the {\smallfonts 2D}
Ising model. Although this result was obtained only up to first
order in the defect amplitude, one may conjecture that it remains
true to all orders, like in the exactly solved limiting cases.

The shape invariance of the perturbation under the special
conformal transformation, when it decays like
a power of the distance to a line, leads on the cylinder to a 
spectrum containing conformal towers. At least in this case 
further work should allow the determination of the 
spectrum--generating algebra.

When the perturbation expansion for the exponents contains divergent
terms, one expects local order at the bulk critical point and a 
first--order defect transition when the couplings are enhanced as
for an internal line defect.
 
Finally let us mention that our results could be extended 
to other two--dimensional conformal systems using conformal
perturbation theory [6, 7].
\ack Usefull collaboration with R. Z. Bariev, F. Igl\'oi and
I. Peschel are gratefully acknowledged.
\references
\numrefjl{[1]}{Cardy J L 1984}{Nucl. Phys. B}{240}{[FS12] 514}
\numrefjl{[2]}{Turban L 1985}{\JPA}{18}{L325}
\numrefjl{[3]}{Guimar\~aes L G and Drugowich de Felicio 
J R 1986}{\JPA}{19}{L341}
\numrefjl{[4]}{Henkel M and Patk\'os A 1987}{\JPA}{20}{2199}
\numrefjl{[5]}{Henkel M, Patk\'os A and Schlottmann M 
1989}{Nucl. Phys. B}{314}{609}
\numrefbk{[6]}{Cardy J L 1987}{Phase Transitions and Critical 
Phenomena}{vol 11 ed. C Domb and J L Lebowitz (London: Academic) 
p 55}
\numrefbk{[7]}{Christe P and Henkel M 1993}{Introduction to 
Conformal Invariance and its Applications to Critical 
Phenomena}{Lecture Notes in Physics (Berlin: Springer) to appear}
\numrefjl{[8]}{Henkel M and Patk\'os A 1987}{Nucl. Phys. 
B}{285}{29}
\numrefjl{[9]}{Henkel M and Patk\'os 1988}{\JPA}{21}{L231}
\numrefjl{[10]}{Baake M, Chaselon P and Schlottmann M 
1989}{Nucl. Phys. B}{314}{625}
\numrefjl{[11]}{Hilhorst H J and van Leeuwen J M J 1981}{Phys.
Rev. Lett.}{47}{1188}
\numrefjl{[12]}{Bl\"ote H W J and Hilhorst H J 1983}{Phys. Rev.
Lett.}{51}{20} 
\numrefjl{[13]}{Burkhardt T W and Guim I 1984}{Phys. Rev.
B}{29}{508}
\numrefjl{[14]}{Burkhardt T W, Guim I, Hilhorst H J and van
Leeuwen J M J 1984}{Phys. Rev. B}{30}{1486}
\numrefjl{[15]}{Peschel I 1984}{Phys. Rev. B}{30}{6783}
\numrefjl{[16]}{Bl\"ote H W J and Hilhorst H J
1985}{\JPA}{18}{3039}
\numrefjl{[17]}{Kaiser C and Peschel I 1989}{J. Stat.
Phys.}{54}{567}
\numrefjl{[18]}{Burkhardt T W and Igl\'oi F 1990}{\JPA}{23}{L633}
\numrefjl{[19]}{Igl\'oi F 1990}{Phys. Rev. Lett.}{64}{3035}
\numrefjl{[20]}{Berche B and Turban L 1990}{\JPA}{23}{3029}
\numrefjl{[21]} {Igl\'oi F, Berche B and Turban L 1990}{Phys. Rev.
Lett.}{65}{1773}
\numrefjl{[22]} {Bariev R Z 1988}{Zh. Eksp. Teor. Fiz.} {94}{374\ 
(1988\ {\frenchspacing\sl Sov. Phys. JETP \bf 67} 2170)}
\numrefjl{[23]} {Bariev R Z 1989}{\JPA}{22}{L397}
\numrefjl{[24]} {Bariev R Z and Malov O A 1989}{Phys. Lett.}{136A}{291}
\numrefjl{[25]} {Bariev R Z and Ilaldinov I Z
1989}{\JPA}{22}{L879}
\numrefjl{[26]}{Bariev R Z and Peschel I 1991}{\JPA}{24}{L87}
\numrefjl{[27]}{Turban L 1991}{Phys. Rev. B}{44}{7051}
\numrefjl{[28]}{Bariev R Z and Peschel I 1991}{Phys.
Lett.}{153A}{166}
\numrefjl{[29]}{Burkhardt T W 1982}{Phys. Rev. Lett}{48}{216}
\numrefjl{[30]}{Cordery R 1982}{Phys. Rev. Lett}{48}{215}
\numrefjl{[31]}{Kramers H A and Wannier G 1941}{Phys. 
Rev.}{60}{252}
\numrefjl{[32]}{Bariev R Z and Turban L 1992}{Phys. Rev.
B}{45}{10761}
\numrefjl{[33]}{Burkhardt T W and Cardy J L 1987}{\JPA}{20}{L233}
\numrefjl{[34]}{Cardy J L 1983}{\JPA}{17}{L385}
\numrefjl{[35]}{Igl\'oi F, Peschel I and Turban L 1993}{Adv.\  Phys.}{42}{683}
\numrefjl{[36]}{Kogut J B 1979}{Rev. Mod. Phys.}{51}{659}
\numrefjl{[37]}{Barber M N, Peschel I and Pearce P A 1984}{J.
Stat. Phys.}{37}{497}
\numrefjl{[38]}{von Gehlen G, Rittenberg V and Ruegg H
1985}{\JPA}{19}{107}
\numrefjl{[39]}{Jordan P and Wigner E 1928}{Z. Phys.}{47}{631}
\numrefjl{[40]}{Lieb E H, Schultz T D and Mattis D C 1961}{Ann.
Phys. (N. Y.)}{16}{406}
\numrefjl{[41]}{Pfeuty P 1970}{Ann. Phys. (N. Y.)}{57}{79}
\numrefjl{[42]}{Peschel I and Wunderling R 1992}{Ann.
Phys.}{1}{125}
\numrefjl{[43]}{Igl\'oi F and Turban L 1993}{Phys.Rev. 
B}{47}{3404}

\vfill\eject\bye